 \documentclass[reprint]{JASA}

\usepackage{algpseudocode}
\usepackage{multirow}
\begin{document}

\title[JASA-Article/LCD inspired from SCD]{Spoken language change detection inspired by speaker change detection}

\author{Jagabandhu Mishra}
\email{jagabandhu.mishra.18@iitdh.ac.in}

\author{S. R. Mahadeva Prasanna}
\email{prasanna@iitdh.ac.in}
\affiliation{Department of Electrical Engineering, IIT Dharwad, Dharwad 580011, Karnataka, India.}

\preprint{Mishra $et ~al.$, LCD inspired from SCD} 	

\date{\today} 

\begin{abstract}

Spoken language change detection (LCD) refers to identifying the language transitions in a code-switched utterance. Similarly, identifying the speaker transitions in a multispeaker utterance is known as speaker change detection (SCD). Since tasks-wise both are similar, the architecture/framework developed for the SCD task may be suitable for the LCD task. Hence, the aim of the present work is to develop LCD systems inspired by SCD.  Initially, both LCD and SCD are performed by humans. The study suggests humans require (a) a larger duration around the change point and (b) language-specific prior exposure, for performing LCD as compared to SCD.  The larger duration requirement is incorporated by increasing the analysis window length of the unsupervised distance-based approach. This leads to a relative performance improvement of $29.1\%$ and $2.4\%$, and a priori language knowledge provides a relative improvement of $31.63\%$ and $14.27\%$ on the synthetic and practical codeswitched datasets, respectively. The performance difference between the practical and synthetic datasets is mostly due to differences in the distribution of the monolingual segment duration.

\end{abstract}


\maketitle
\section{\label{sec:1} Introduction}

Spoken language diarization (LD) is a task to automatically segment and label the monolingual segments in a given multilingual speech signal. The existing works towards  LD are very few~\cite{surveycodeswitch2019}. The majority of them use phonotactic (i.e. the distribution of sound units) based approaches~\cite{chan2004detection,lyu2013language, V2018}. The development of LD  using a phonotactic-based approach requires transcribed speech utterances. The same is difficult to obtain as most of the  languages present in the code-switched multilingual utterances are resource-scare in nature~\cite{surveycodeswitch2019, V2018}. Even though, there exist some transfer learning approaches that adapt the phonotactic models of the high resource language to obtain the models for the low resource language, may end up with performance degradation if both the languages are not from the same language group~\cite{surveycodeswitch2019}. Further, LD is effortless for humans, especially for known languages, and challenging for machines. Hence there is a need for exploring alternative approaches for LD. 

Speaker diarization (SD) is a task to automatically segment and label the mono-speaker segments for a given multispeaker utterance, which is well explored in the literature. Though there exist differences in the information that needs to be captured to perform LD and SD tasks, there exist many similarities like the features approximating the vocal tract resonances that have been successfully used for the modeling of both speaker and language-specific phonemes~\cite{carrasquillo2002approaches,li2013spoken,liu2021end}.  Furthermore, most of the approaches used for spoken language identification (LID) are inspired by the approaches used for the speaker identification/verification (SID/SV) task~\cite{snyder2018spoken,richardson2015deep}. In addition to that most of the successful LID systems that are borrowed from SID/SV literature do not require transcribed speech data~\cite{li2013spoken,snyder2018spoken}. Alternatively LID systems developed using the phonotactic approach require transcribed speech data. This motivates a close association study between the LD and SD tasks and may be exploited to come up with approaches for LD.  

The SD field has evolved mainly in two ways: (1) change point detection followed by clustering and boundary refinement, and (2) fixed duration segmentation followed by i-vector/  embedding vector extraction, clustering, and boundary refinement~\cite{park2022review,moattar2012review,tranter2006overview}. ~\cite{hemaSD2021,bredin2017speaker,hogg2019speaker,park2022review} reported that initial change point detection improved overall SD performance. Thus this study focuses on the development of spoken language change detection (LCD)  through a comparative analysis between LCD and speaker change detection (SCD). The available SCD approaches can be broadly classified into two groups: (1) distance-based unsupervised approach and (2) model-based supervised approach~\cite{moattar2012review,park2022review}. The distance-based approach applies hypothesis testing (either coming from a unique speaker or not) for predicting the speaker change to the speaker's specific features extracted from the speech signal with sliding consecutive windows~\cite{moattar2012review,park2022review}. Following this approach, many feature extraction techniques like excitation source~\cite{dhananjaya2008speaker,sarma2015speaker}, fundamental frequency contour~\cite{hogg2019speaker}, etc., and distance metrics like Kullback–Leibler (KL) divergence~\cite{siegler1997automatic}, Bayesian information criteria (BIC)~\cite{chen1998speaker}, KL2~\cite{siegler1997automatic}, generalized likelihood ratio (GLR)~\cite{gish1991segregation} and information bottleneck (IB)~\cite{hemaSD2021} are proposed in the literature. Generally, the performance of the distance-based unsupervised approach degrades with variation in environment and background noise (it may predict false changes), hence to resolve the issue supervised model-based approaches are proposed in the literature~\cite{moattar2012review,park2022review}. In the early days, the proposed approaches model individual speakers using the Gaussian mixture model and universal background model (GMM-UBM)~\cite{barras2006multistage,moattar2012review}, hidden Markov model (HMM)~\cite{meignier2006step}, etc, but nowadays, using the deep learning framework the approach predicts the speaker change by discriminating between the speaker change segments (neighborhood of the speaker change point) with no change segments~\cite{moattar2012review,park2022review}. However, the model-based approach smooths the output evidence and may lead to miss detection of the change points~\cite{moattar2012review}. In addition to that training of the supervised model requires labeled speech data from a similar environment/recording condition, speaking style, language, etc., making the system development complicated. Therefore the distance-based unsupervised approaches are more popular and widely used for SCD tasks~\cite{hemaSD2021,park2022review,moattar2012review}.

\begin{figure}
\includegraphics[height= 160pt,width=240pt]{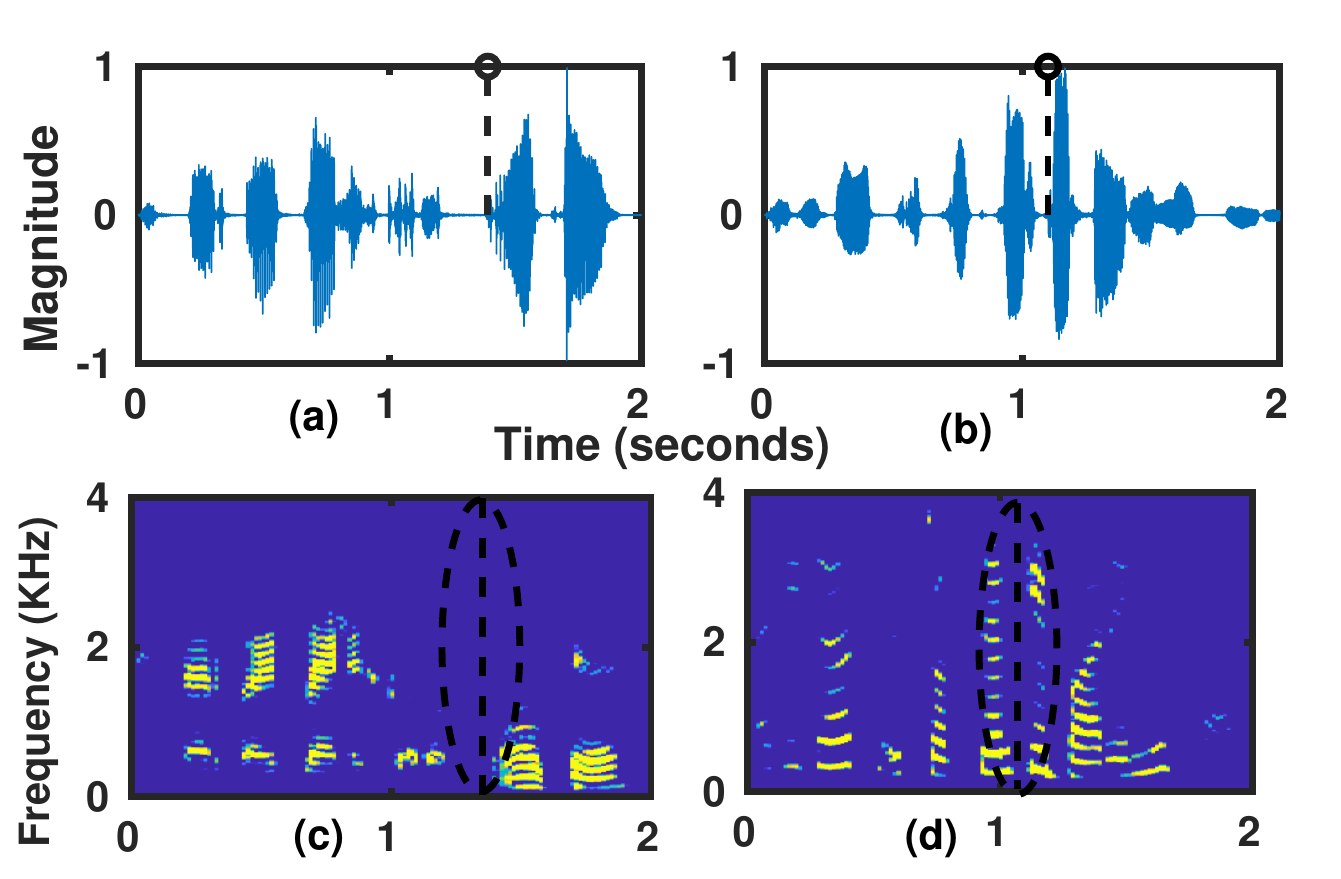}
 \caption{(a) and (c) Two speaker time domain speech signal  and its spectrogram, respectively. (b) and (d) Two languages (Bilingual) time domain speech signal and its spectrogram, respectively.}
 \label{spk_ln_ch}
\end{figure}

Even though the available SCD frameworks look simple to adopt, there are challenges in doing so. Fig.~\ref{spk_ln_ch} (a) and (b), show the time domain speech signals corresponding to the utterance having a speaker change and a language change, respectively. By listening and observing the time domain representation of both utterances, the identified speaker/language change points are manually marked. From the time domain signal, it is very difficult to locate both the speaker and language change points. Fig.~\ref{spk_ln_ch} (c) and (d) show the spectrogram  of both utterances. From the spectrogram, it can be observed that around the speaker change the formant structure shows significant variation, whereas around language change the structure is intact. When the speaker changes, the vocal tract system information changes and hence the variation in the formant structure. However, the structure of the formant frequencies remains intact during language change as the single speaker is speaking both languages. It is interesting to note that humans discriminate between spoken languages without knowing the detailed lexical rules and phonemic distribution of the respective languages. Of course, humans need to have prior exposure to the languages~\cite{li2013spoken}. Humans  may exploit the long-term phoneme dynamics to discriminate between languages. Therefore, the language change may be detected by capturing the long-term language-specific spectral-temporal dynamics. This may represent valid phoneme sequences and their combinations to form syllables and subwords of a language.

Based on the need to exploit the long-term spectro-temporal evidence, it can be hypothesized that the LCD by human/machine may require more neighborhood duration around the change point than the SCD. In addition, LCD may also benefit from prior exposure to respective languages. A human subjective study that focuses on language/speaker change detection is set up for validating the same. 

For automatic detection of language change, the initial studies are performed using the available unsupervised distance and the supervised model-based SCD approaches. The model-based approaches include GMM-UBM, i-vector, and x-vector. Based on the experimental results for LCD and SCD, appropriate modifications will be done to each framework for improving the performance of the LCD task.


The main contribution of this work are summarized as follows:  (a) by observing the spectro-temporal representation around the speaker and language change, it is hypothesized that detecting language change, requires a larger duration around the change point and a priori knowledge of the language as compared to detecting a speaker change. The same hypothesis is confirmed by the human subjective study, (b) the SCD frameworks are used as initial baselines to perform LCD and their performances are analyzed, and (c) these frameworks are further refined to improve the performance of LCD.


\section{Database setup}

This section provides a brief description of the database used in this study. For performing the LCD/SCD task among humans, we have selected $32$ and $15$ utterances for the language and speaker change study, respectively. All the utterances have only one language/speaker change point and have approximately $6-8$ syllables on either side of the change point. For the language change study, we have selected $32$ utterances from the publicly available sources (mostly from Youtube), whereas we have chosen $15$ utterances from the IITG-MV phase 3 and DIHARD datasets for the speaker change study~\cite{haris2012multivariability,ryant2018first}. The $32$ utterances used for the language change study are from the $10$ language pairs and have $4,4,4,4,4,2,2,4,2,2$ utterances, respectively from, (1) Hindi-English (HIE), (2) Bengali-English (BEE), (3) Telugu-English (TEE), (4) Tamil-English (TAE), (5) Bengali-Assamese (BEA), (6) Bengali-Bengali (BEB), (7) Assamese-Assamese (ASA), (8) Tamil-Malayalam (TAM), (9) Tamil-Tamil (TAT) and, (10) Malayalam-Malayalam (MAM), respectively. It is difficult to get the utterances having language pairs, Bengali-Assamese, and Tamil-Malayalam spoken by a single speaker. Hence, these language pairs with and without having a language change are considered along with a speaker change. The selected utterances for both LCD and SCD tasks along with their change point annotations are available at~\href{https://github.com/jagabandhumishra/HUMAN-SUBJECTIVE-STUDY-FOR-LCD-and-SCD} {\url{https://github.com/jagabandhumishra/HUMAN-SUBJECTIVE-STUDY-FOR-LCD-and-SCD}}.

Initially, the studies have been performed with synthetically generated code-switch and multi-speaker utterances. For generating the utterances, we have used the Indian institute of technology Madras text-to-speech (IITM-TTS) corpus ~\cite{baby2016resources}. The IITM-TTS  corpus consists of speech data recordings from native speakers of $13$ Indian languages. For each native language, two speakers (a male and a female) recorded their utterances in their native language and English. In this study for synthesizing the code-switch utterances, a female speaker speaking her native language Hindi, and her second language English is considered. For each language, the first $5$ hours of data are used for training purposes. The rest of the monolingual utterances are stitched randomly for generating code-switched utterances. Altogether, $4000$ utterances are generated having one to five language change points. The average monolingual segment duration of the generated code-switch utterances for Hindi and English languages are approximately $6.5$ and $5.2$ secs,  respectively. The generated dataset is termed as TTS female language change (TTSF-LC) corpus. Similarly, for generating speaker change utterances by keeping the language identical, we have used English speech utterances from native Hindi and Assamese female speakers. The average mono-speaker segment duration of the generated utterances are $5.19$ and $4.86$ secs, respectively. The generated dataset is termed as TTS female speaker change corpus (TTSF-SC).


Finally, for generalizing the obtained observations, the experiments are performed on the standard LCD corpus. Microsoft code-switched challenge task-B (MSCSTB) dataset is used. The dataset has development and training partitions that consist of code-switched utterances and language tags (each $200$ msec) from three language pairs: Gujarati-English (GUE), Tamil-English (TAE), and Telugu-English (TEE). The approximate duration of each language in the training and development set is $16$  and $2$ hours, respectively. The detail about the database can be found at~\cite{diwan2021multilingual}. 



\section{Human subjective study for language and speaker change detection}
\label{sub_study}

An experimental procedure has been set up, where each human subject is exposed to a pool of utterances that may or may not have a language/speaker change. The human subjects are asked to mark, if there exists a language/speaker change or not. The utterances are classified into five groups. Each group is represented with approximate duration considered in terms of the number of voiced frames (NVF) taken around the true/false change point. The true change point refers to the actual change points of the selected utterances. The selected utterances are split around the change point to generate the mono-language/speaker utterance. The false change point represents the centered voiced frame's start location of the given mono-language/speaker utterance. The voiced frame is decided by taking $6\%$ of the average short time frame energy (computed with a frame size of $20$ msec and a frameshift of $10$ msec) of a given utterance as a threshold~\cite{rabiner1978digital}. The $30$ mono-speaker utterances are generated by splitting the selected $15$ utterances around the true change point. Out of $30$, with respect to duration, the largest $15$ has been chosen for this study. The same procedure has been followed to generate the mono-lingual utterances using the selected code-switched utterances belonging to the HIE, BEE, TAE, and TEE language pairs. However, there is an exception for the utterances belonging to BEA and TAM, as the utterances have a speaker change along with the language change. Hence for a fair comparison, the mono-lingual utterances for these cases are synthesized, such that they also have a speaker change, i.e. BEB, ASA, MAM, and TAT, respectively. After that, each utterance $S(n)$ is masked by considering $x$ number of voiced frames (NVF-$x$) from the left and right of the true/false change point. According to the value of $x$, the masked utterances are grouped into five different groups, termed NVF-$10$, NVF-$20$, NVF-$30$, NVF-$50$, and NVF-$75$. To avoid abrupt masking, a Gaussian mask $G(n)$ with appropriate parameters is multiplied with the utterances to obtain the masked utterance $S_{m}(n)=S(n) \times G(n)$. The masked signal is passed through an energy-based endpoint detection algorithm to obtain the final masked utterance~\cite{rabiner1978digital}. The detailed procedure of the masked utterance generation is attached in the supplementary\footnote{Supplementary material for the data generation procedure for the LCD and SCD study among humans is available at [AIP will insert URL]},  and also the generated utterances are available at ~\href{https://github.com/jagabandhumishra/HUMAN-SUBJECTIVE-STUDY-FOR-LCD-and-SCD} {\url{https://github.com/jagabandhumishra/HUMAN-SUBJECTIVE-STUDY-FOR-LCD-and-SCD}}.

The listening experiment is conducted with $18$ subjects. Out of them, $13$ number of the subjects are male and $5$ are female. The selected subjects are from the $20-30$ years age group. The subjects have no prior exposure to the voice samples of the speakers used in this study. However, the subjects are comfortable with English, and for other languages, the comfortability varies. To know the language comfortability, each of the subjects is asked to provide a language comfortability score (LCS) from zero to three for each pair of languages. 

The listening study is conducted with $390$ utterances (i.e $240$ for LCD and $150$ for SCD). The LCD task is separate from SCD, hence conducted in two different sessions, and also the subjects are well rested so that they don't have listener fatigue. A graphical user interface (GUI) has been designed to perform the listening study. For a specific LCD/SCD study, all the masked utterances are presented to the listener in a random order, irrespective of their segment duration.  If a listener is unable to provide the response for one-time playing,  s/he is allowed to play the utterance multiple times. Our objective here is to observe, how correctly humans recognize the speaker and language change by listening to the utterances coming from the five different groups. Hence, the responses recorded in ~\cite{sharma2019talker} for analyzing the talker change detection ability of humans are used here. Three kinds of responses have been recorded, these are (1) language/speaker change detected or not (2) the number of times replayed (NR), and (3) response time (RT). RT is the time duration taken by a subject to provide his/her response, after listening to the full utterance. The RT is computed by subtracting the respective utterance duration (UD) from the total duration (TD) (i.e. $RT=TD-UD$). The TD is the duration taken by a subject (i.e. from pressing the play button to pressing the yes/no button) to provide his/her response.

\begin{equation}
\label{der}
    DER=\frac{(FA+FR)}{N}
\end{equation}

For a given subject, there are three kinds of performance measures computed in this study: (1) average detection error rate ($DER$) (2) average number of times replayed ($NR$),  and (3) average response time ($RT$). The $DER$ is defined in Eq.~\ref{der}, where $N$ is the total number of trials, $FA$ is the number of false language/speaker change utterances, marked as true by the subject and $FR$ is the number of true language/speaker change utterances, marked as false by the subject, respectively. The $DER$ measure defines the inability of the subject to detect language/speaker change. The $NR$ provides an estimation of the average number of replays required for the subject to mark their response comfortably. Similarly, the $RT$ provides an estimation of the average duration required for the subject to perceive the language/speaker change, after listening to the respective utterances. A higher value of the performance measures indicates the inability of the human subject to perceive the language/speaker change and vice versa.

\begin{figure} 
\includegraphics[height= 200pt,width= 240pt]{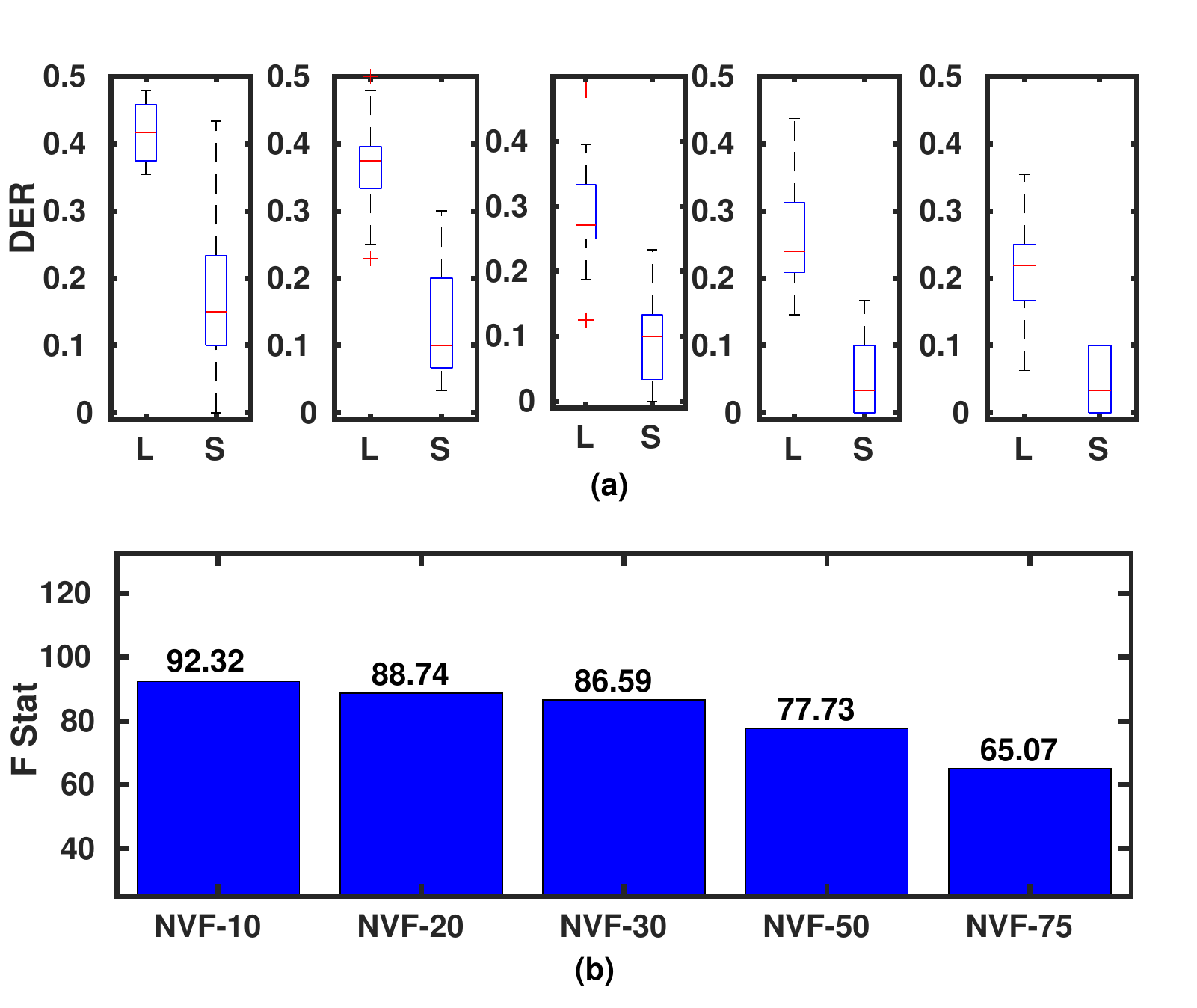}
\caption{(a) $DER$ distributions of the subjects,  (b) F-Statistics (F Stat) values of the ANOVA test between the $DER$ distributions of LCD (L) and SCD (S) study, respectively .}
 \label{sub_res_1}
\end{figure}

\begin{figure} 
\includegraphics[height= 200pt,width= 240pt]{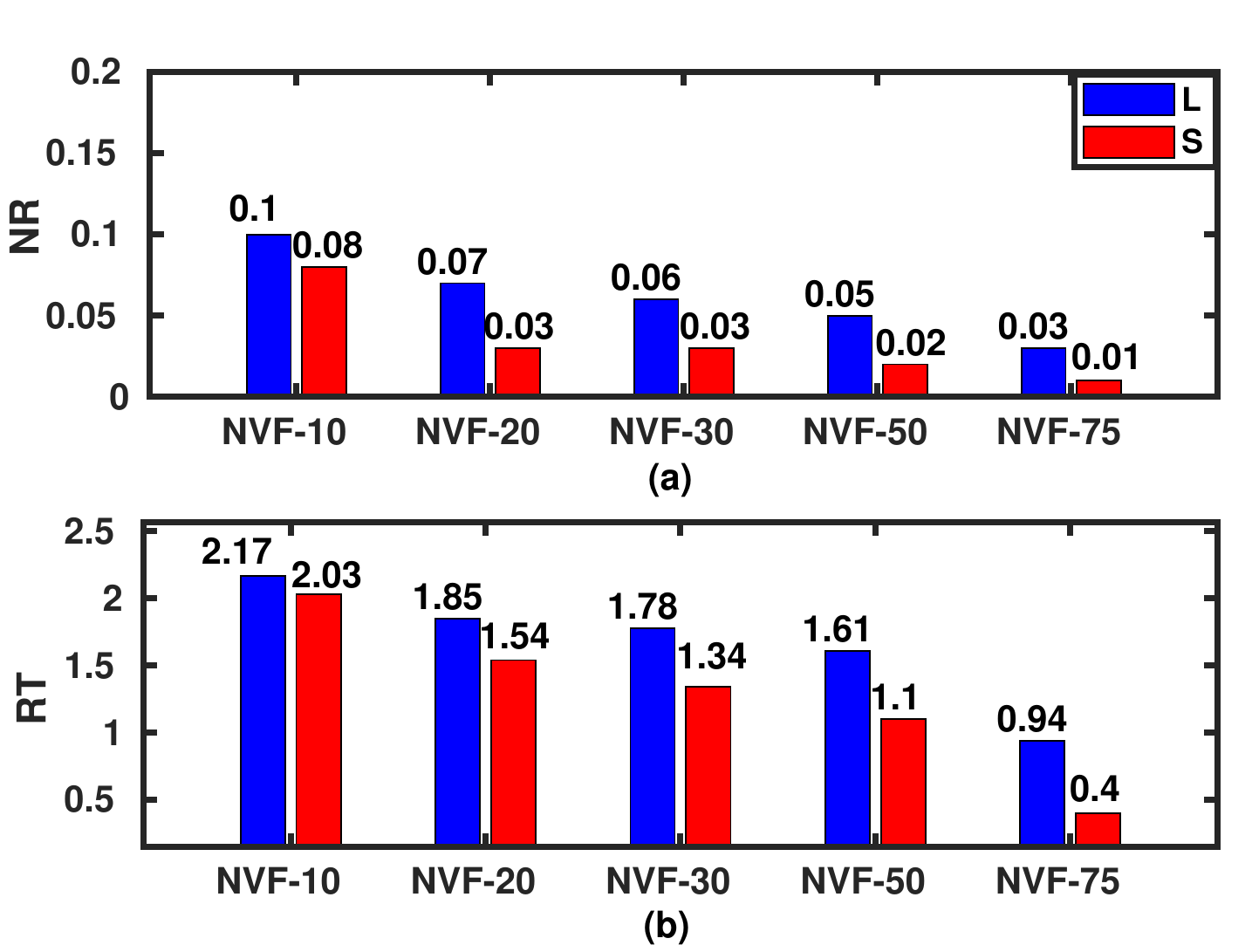}
\caption{ Median values of the (a) $NR$ and (b) $RT$ distributions for the LCD and SCD.}
 \label{sub_res_2}
\end{figure}

After performing both the LCD and SCD experiments, the subject-specific, $DER$, $NR$, and $RT$ are computed with respect to NVF. The distributions of the obtained $DER$ with respect to the NVF are depicted in Fig.~\ref{sub_res_1}(a). It can be seen that the $DER$ values are smaller for the SCD than for the LCD, regardless of the NVF. This suggests that human subjects are more comfortable with detecting the switching of speakers than language. Furthermore, as the NVF increases from $10$ to $75$, the DER decreases for both SCD and LCD.  The differences between the DER distribution of the LCD and SCD decrease with an increase in the NVF. This suggests that human subjects' comfortability in detecting language change increases and becomes at par with speaker change, with the increase in the NVF. To further validate fact, a statistical test called an analysis of variance (ANOVA) has been performed between the $DER$ distribution (after removing the outliers) of LCD and SCD. The obtained F-statistics values are depicted in Fig.~\ref{sub_res_1}(b). The higher F-statistics value suggests having better discrimination between the two distributions and vice-versa. From the figure, it can be observed the F-statistics values reduced with an increase in NVF. This justifies the claim that humans' language discrimination ability improves and goes closure to the speaker discrimination ability with an increase in NVF.  The median values of the recorded $NR$ and $RT$ values are depicted in Fig.~\ref{sub_res_2}. It can be observed from the figure that, like $DER$, the median value of $NR$, and $RT$ reduces with an increase in NVF. The median values of $NR$ and $RT$ are also smaller for SCD than LCD. This concludes that human subjects require a larger duration around the change point to detect language than the speaker change comfortably.

\begin{figure} 
\includegraphics[height= 150pt,width= 200pt]{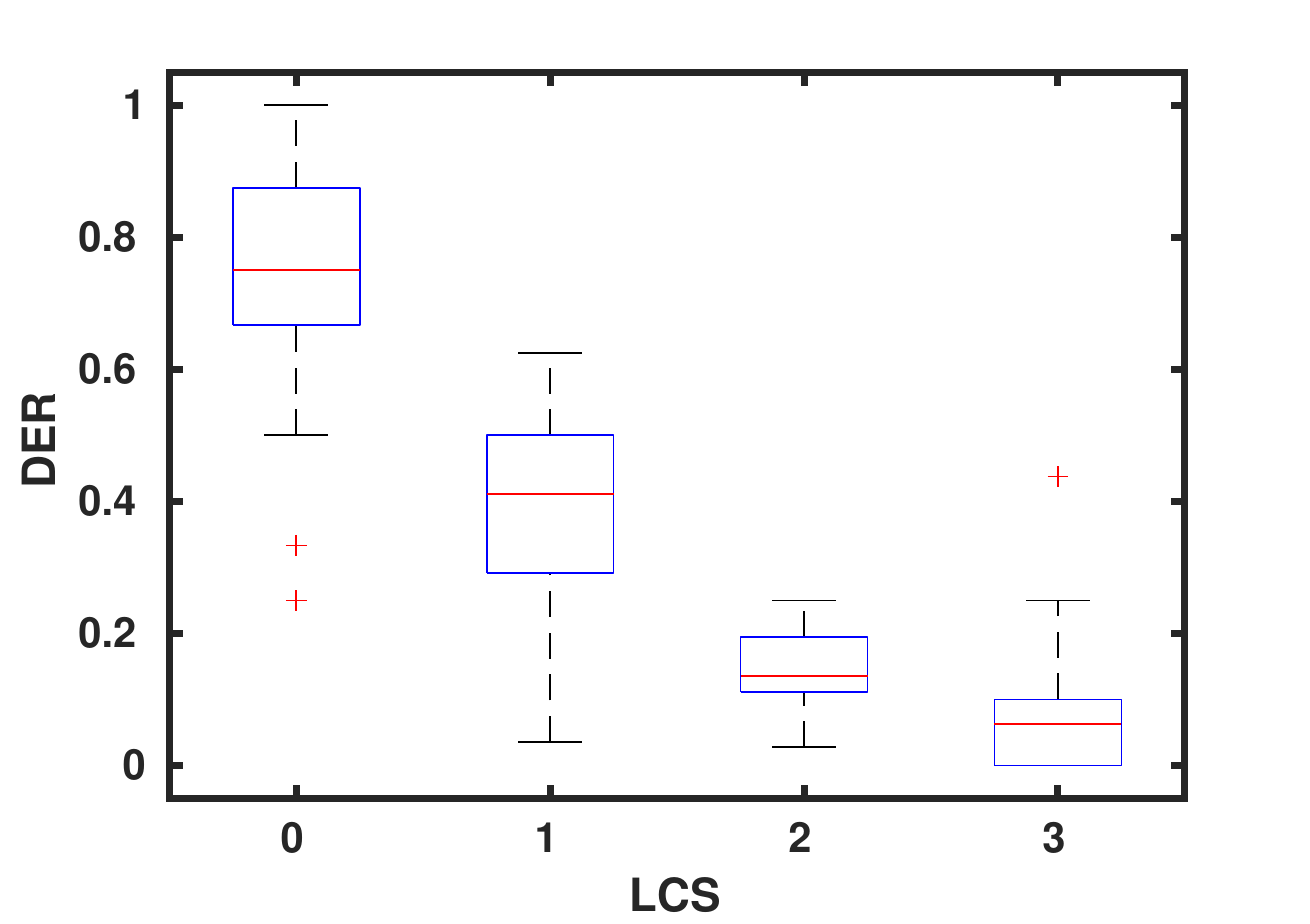}
\caption{DER vs. language comfortability score (LCS) for LCD with NVF-50 and NVF-75. }
 \label{sub_res_3}
\end{figure}

For observing the effect of language comfortability on detecting language change, the responses of the human subjects are considered for the group NVF-$50$ and NVF-$75$ that have the median of $DER$ lesser than $0.25$ (assuming sufficient duration from either side). With respect to the LCS, the responses are segregated into four groups. The group segregation with respect to language comfortability is done as $0$: very low, $1$: lower medium, $2$: medium, and $3$: excellent, respectively. The obtained $DER$ distribution with respect to LCS is depicted in Fig.~\ref{sub_res_3}. From the figure, it can be observed that the $DER$ values are decreases with an increase in LCS. This concludes that a priori knowledge of languages helps people to better discriminate between languages.


\section{LCD and SCD using unsupervised distance-based approach}
\label{ob_study}

\begin{figure} 
\includegraphics[height= 140pt,width= 230pt]{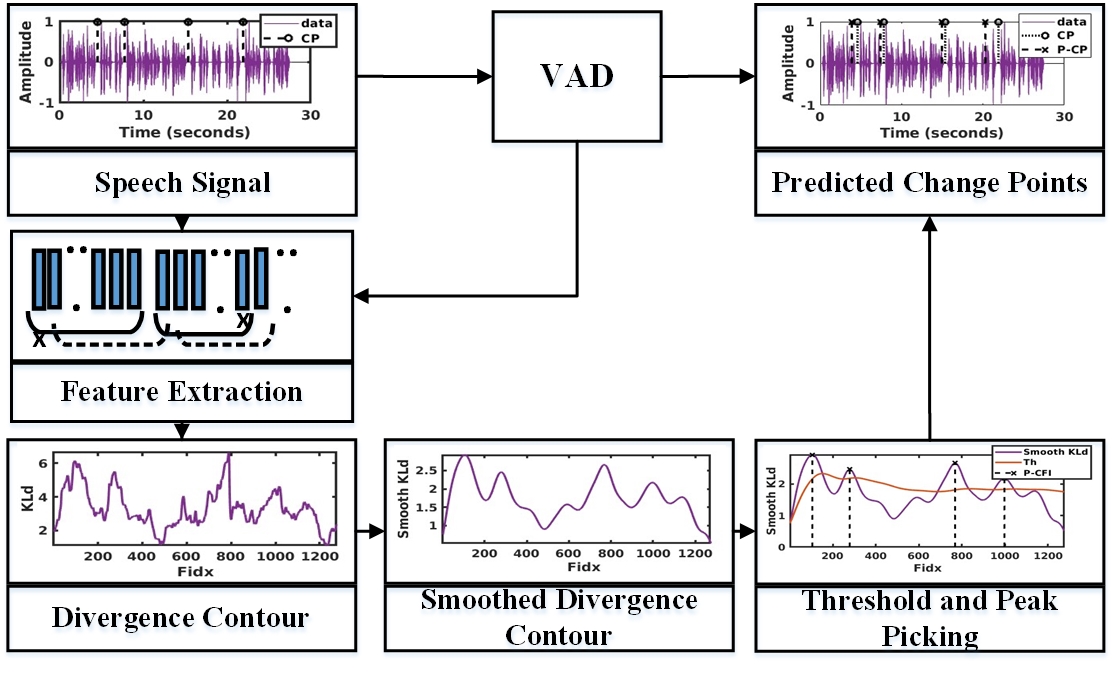}
\caption{Basic block diagram of the change detection framework for unsupervised distance-based approach}
 \label{cdp}
\end{figure}

The objective of this section is to perform LCD tasks inspired by the existing unsupervised distance-based SCD framework. In general, the SCD task is performed by computing and threshold the distance contour obtained between the features of the sliding analysis window with a fixed length $N$. The basic block diagram of the approach is depicted in Fig.~\ref{cdp}. First feature vectors are extracted from the speech signal and then energy-based voice activity detection (VAD) is performed to obtain the voiced frame indices. The voiced frame indices are stored for future reference and the feature vectors corresponding to the voiced frames are used for further processing. The voiced feature vectors are used with two consecutive windows having a fixed length to model two different Gaussian distributions ($g_{a}$ and $g_{b}$). The divergence distance contour is obtained through the entire scan of the given test utterance by sliding the analysis window with a frame, as mentioned in Eq.~\ref{kld}. The evidence contour is then smoothed with the hamming window with length ($h_{l}$). The smoothed contour is then used for peak detection, with a peak-picking algorithm having a minimum peak distance parameter called $\gamma$. The higher value of $\gamma$ reduces the number of detected peaks and vice-versa. For reducing the number of false change points, an approach of deriving a threshold counter proposed in~\cite{lu2002speaker} and mentioned in Eq~\ref{kld_th} is used here. Finally, the change frame is obtained by comparing the strength of the detected peaks with the threshold contour. The change point's actual frame index and sample location are obtained by using the stored voiced frame locations.

\begin{equation}
\label{kld}
  D(i)=KL(g_{A}|g_{B})+KL(g_{B}|g_{A}),\\
\end{equation}

\begin{equation}
\label{kld_th}
 Th(i) =\alpha.\frac{1}{N}\sum_{n=0}^{N}D(i-n-1,i-n)
\end{equation}

Initially, we used the TTSF-SC dataset for designing and tuning the hyperparameters of the SCD system.  Out of $4000$ test utterances, the first $100$ utterances are used to tune the hyperparameters. It has been observed that the performance is optimal by considering $\alpha=1$, $\gamma$ equal to $0.9$ times the analysis window length, and $150$ as the analysis window length.  Keeping the methodology and hyperparameters identical, the TTSF-LC and MSCSTB dataset is used to perform the LCD task. For evaluating the performance, the commonly used performance measures for event detection tasks, i.e. identification rate (IDR), false acceptance rate (FAR), miss rate (MR), and mean deviation ($D_{m}$) are used here~\cite{murty2008epoch,mishra2021spoken}. The performances of both tasks are tabulated in Table~\ref{dp}.

From the results, it can be observed that the performance of the SCD in terms of IDR is $84.1\%$, whereas the performance of the LCD in terms of IDR is $51.2\%$. The reduction in performance may be due to two reasons, (1) the used MFCC features may fail to capture language-specific discriminative evidence, and (2) the hyperparameters, mostly the analysis window length, are tuned for SCD and may not be appropriate for LCD. Hence to understand the issue a study is carried out by varying the features and analysis window length around the change point. The most used features in literature for language identification (LID) tasks, i.e. MFCC, LPCC, SDC, and PLP are considered here. The objective here is to observe the language discriminative ability of the features by considering a fixed number of voiced frames (NVF),  $x$ around the change point and compare it with the speaker discrimination ability of the MFCC feature. This study will help us to reason out the performance degradation of LCD as compared to SCD. Further, the observation will also help us to optimally decide the feature and analysis window length for performing LCD.

\begin{figure}
\includegraphics[height= 120pt,width= 230pt]{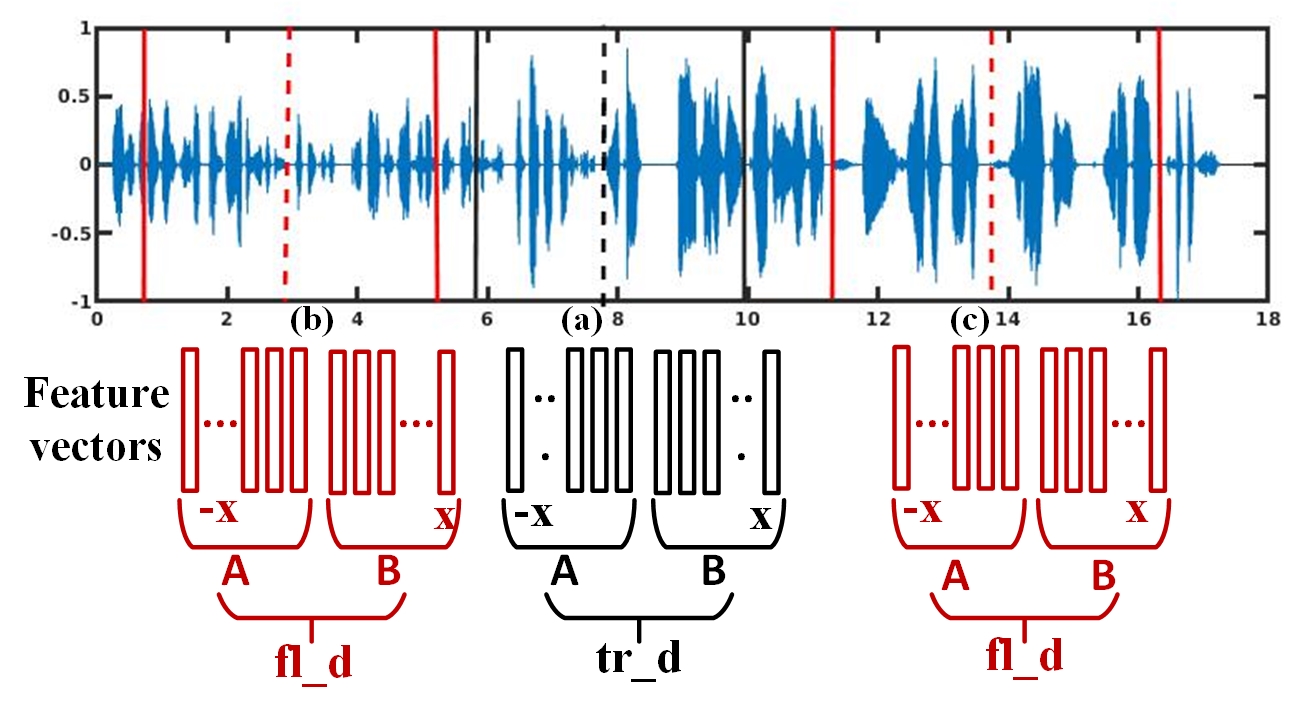}
\caption{Distance computation around the true and false change point of an utterance, (a) true change point  and (b), (c) false change points, fl\_d, and tr\_d are false and true distances, respectively. }
 \label{seg_gen}
\end{figure}

\begin{figure} 
\includegraphics[height= 130pt,width= 230pt]{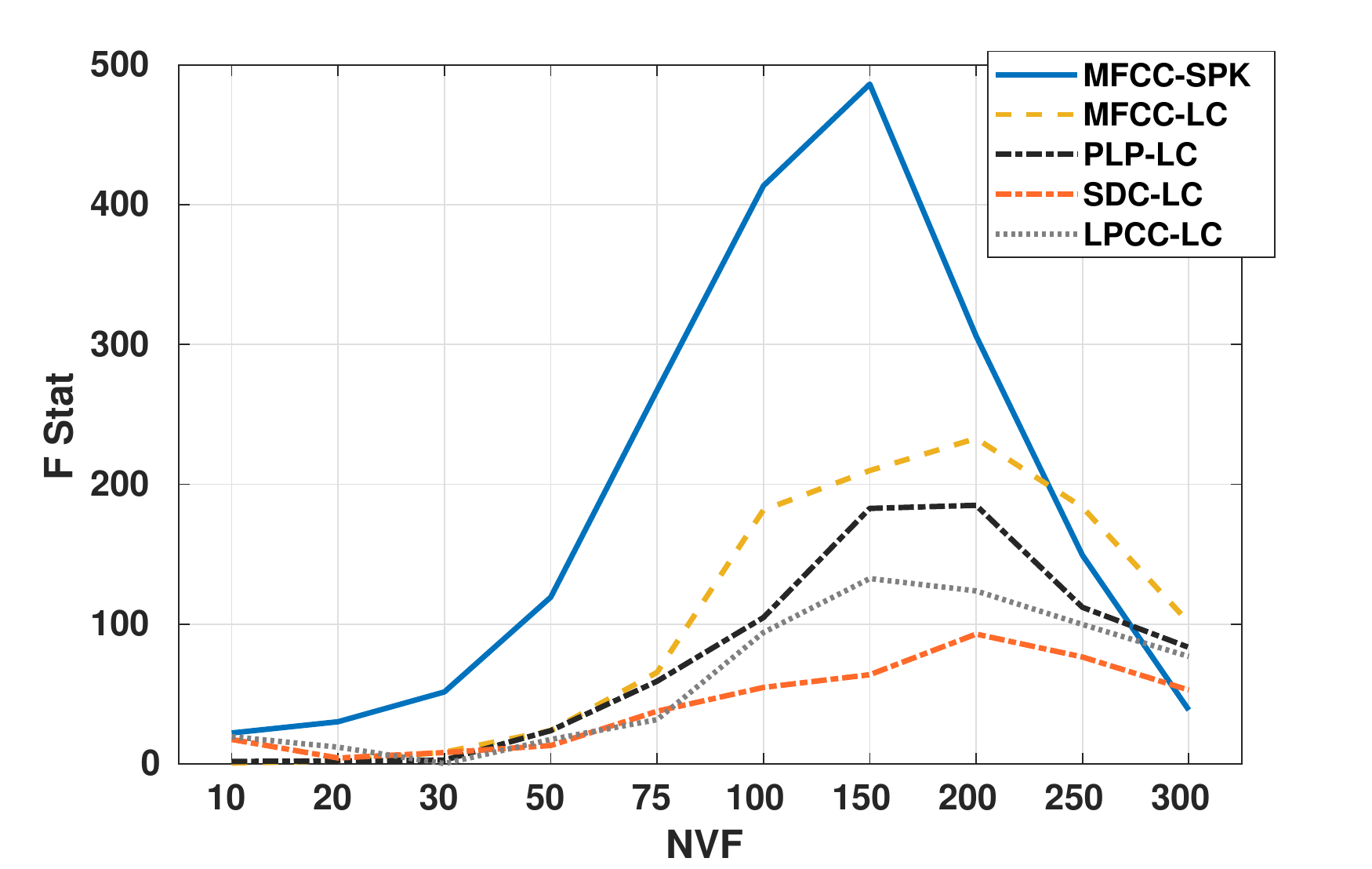}
\caption{ANOVA test F-statistics (F Stat) values obtained between the true and false KL divergence distances for speaker/language change study with varying the number of voiced frames (NVF).}
 \label{fs_anova}
\end{figure}

For performing the study, the 	TTSF-SC and TTSF-LC dataset is considered. Out of $4000$ test utterances, the utterances having only one change point are selected. The number of utterances selected for speaker change and language change is $799$ and $836$, respectively. For observing the discrimination ability, the idea here is to observe the distributional difference between the true and false distances. The true distances are the KL divergence distance between the $x$ number feature vectors from either side of the ground truth change point. Similarly, the false distance is computed by placing the change point randomly anywhere in the mono-language/ speaker segments. The procedure of computing the true and false distances is also depicted in Fig.~\ref{seg_gen}. For observing the duration effect on the discrimination, the value $x$ is considered as $10$, $20$, $30$, $50$, $75$, $100$, $150$, $200$, $250$, and $300$, respectively. For a given $x$ value, the ANOVA test is conducted between the obtained true and false distances. The obtained F-statistics values of the ANOVA test are depicted in Fig.~\ref{fs_anova}.  

\begin{figure} 
\includegraphics[height= 160pt,width= 230pt]{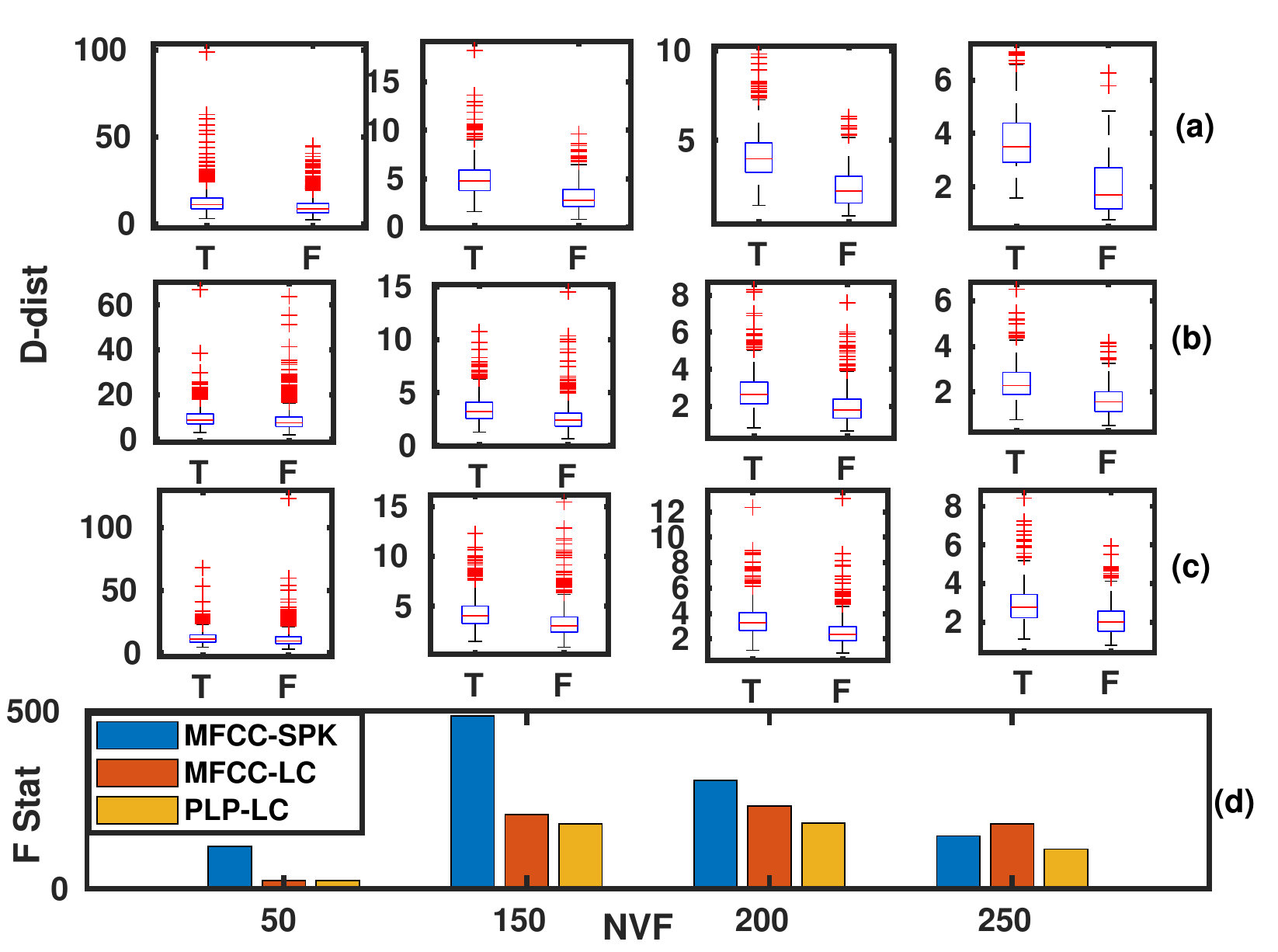}
\caption{True (T) and false (F) distance distribution (D-dist) (a) with MFCC feature for speaker change, (b) with MFCC feature for language change, (c) with PLP feature for language change, and (d) shows the corresponding F-statistics (F Stat) values.}
 \label{box_d}
\end{figure} 

\begin{table*}[]
\centering
\caption{Perforance of LCD and SCD with the unsupervised distance-based approach. A: with $N=150$ (tuned for SCD) and B: with the optimal $N$ value (tuned for LCD).}
\label{dp}
\begin{tabular}{|l|c|ccc|cccccc|}
\hline
\multirow{4}{*}{} &
  \multirow{2}{*}{TTSF-SC} &
  \multicolumn{3}{c|}{\multirow{2}{*}{TTSF-LC}} &
  \multicolumn{6}{c|}{MSCSTB} \\ \cline{6-11} 
 &
   &
  \multicolumn{3}{c|}{} &
  \multicolumn{2}{c|}{GUE} &
  \multicolumn{2}{c|}{TAE} &
  \multicolumn{2}{c|}{TEE} \\ \cline{2-11} 
 &
  MFCC &
  \multicolumn{2}{c|}{MFCC} &
  PLP &
  \multicolumn{6}{c|}{MFCC} \\ \cline{2-11} 
 &
  A &
  \multicolumn{1}{c|}{A} &
  \multicolumn{1}{c|}{B} &
  B &
  \multicolumn{1}{c|}{A} &
  \multicolumn{1}{c|}{B} &
  \multicolumn{1}{c|}{A} &
  \multicolumn{1}{c|}{B} &
  \multicolumn{1}{c|}{A} &
  B \\ \hline
\multicolumn{1}{|c|}{IDR} &
  84.1 &
  \multicolumn{1}{c|}{51.2} &
  \multicolumn{1}{c|}{66.1} &
  64.06 &
  \multicolumn{1}{c|}{42.9} &
  \multicolumn{1}{c|}{44.07} &
  \multicolumn{1}{c|}{47.4} &
  \multicolumn{1}{c|}{48.75} &
  \multicolumn{1}{c|}{44.5} &
  45.24 \\ \hline
\multicolumn{1}{|c|}{FAR} &
  10.7 &
  \multicolumn{1}{c|}{41.3} &
  \multicolumn{1}{c|}{20.91} &
  22.58 &
  \multicolumn{1}{c|}{8.1} &
  \multicolumn{1}{c|}{9.24} &
  \multicolumn{1}{c|}{8.2} &
  \multicolumn{1}{c|}{8.57} &
  \multicolumn{1}{c|}{7.7} &
  7.70 \\ \hline
\multicolumn{1}{|c|}{MDR} &
  5.2 &
  \multicolumn{1}{c|}{7.5} &
  \multicolumn{1}{c|}{12.98} &
  13.36 &
  \multicolumn{1}{c|}{49} &
  \multicolumn{1}{c|}{46.69} &
  \multicolumn{1}{c|}{44.4} &
  \multicolumn{1}{c|}{42.68} &
  \multicolumn{1}{c|}{47.8} &
  47.06 \\ \hline
\multicolumn{1}{|c|}{Dm} &
  0.19 &
  \multicolumn{1}{c|}{0.45} &
  \multicolumn{1}{c|}{0.51} &
  0.57 &
  \multicolumn{1}{c|}{0.5} &
  \multicolumn{1}{c|}{0.49} &
  \multicolumn{1}{c|}{0.5} &
  \multicolumn{1}{c|}{0.56} &
  \multicolumn{1}{c|}{0.5} &
  0.56 \\ \hline
\end{tabular}
\end{table*}

From the figure, it can be observed that the F-statistics values increase with an increase in NVF and saturate after a certain number of voiced frames, and started decreasing after that. A similar observation has also been observed in the case of the LCD and SCD study by humans. However, in case humans' performance doesn’t degrade with an increase in NVF. This may be due to the inability of the Gaussian (assumption of statistical independence) to model the speaker and language spectral dynamics and leading to the increase of the class-specific variance in the distance distribution. Using the MFCC feature, the F-statistic values of the SCD are higher than the LCD irrespective of the NVF. Further, it can also be observed that the discrimination ability (in terms of F-statistics) of the LCD follows the SCD with an increase in the NVF. Furthermore, it has also been observed that the highest F-statistics values obtained for speaker and language change study are at $150$ and $200$, respectively. 

In addition to this,  for language change study, the MFCC features provide better F-statistics value, followed by PLP, LPCC, and SDC. For clear observation, the distance distribution of the MFCC feature to perform SCD and the MFCC and PLP features to perform LCD with NVF of $50$, $150$, $200$, and $250$ is depicted through box plots in Fig.~\ref{box_d}. From the box plots, it can also be noticed that the speaker and language discrimination saturates at NVF $150$ and $200$, respectively. Though the boxplots look to have better discrimination, the increase in inter-class variance leads to a decrease of the F-statistics values. Furthermore, the discrimination ability of the MFCC is better compared to PLP, as the separation between the true and false distance distribution of the MFCC feature is higher than the PLP feature for LCD at NVF equal to 200. This motivates us to consider the MFCC feature with the analysis window length of $200$ for performing LCD for the TTSF-LC dataset. The performance of the LCD task with modified analysis window length is tabulated in Table~\ref{dp}.

The table shows that the performance in terms of IDR, FAR, and MDR follows the observations noticed with respect to the F-statistics. The performance obtained for the TTSF-LC dataset with MFCC feature (considering analysis window length $200$) is $66.1\%$ in terms of IDR, providing a  relative improvement of $29.1\%$ and followed by the IDR of $64.06\%$ using PLP feature. Similar observations also have been reported using the MSCSTB dataset, where the performance in terms of IDR improved relatively with $2.72\%$, $2.85\%$, and $1.63\%$ by considering the analysis window length of $160$, $180$, and $170$ for GUE, TAE, and TEE language pairs, respectively. The analysis window length $160$, $180$, and $170$ are decided greedily by evaluating the performance by considering the analysis window length from $100$  to $250$ with a shift of $10$ on the first $100$ test trails. Hence, this justifies the hypothesis that the requirement of relatively higher duration information to perform LCD than SCD.

\section{Language Change Detection by Model-based Approach}
\label{model_exp}
The SCD and LCD by human suggest that prior exposure to the language make human more efficient in detecting language change.  This motivates extracting the statistical/embedding vectors from the trained machine learning (ML)/ Deep learning (DL) framework and using them to perform change detection tasks. The detailed procedure is explained in the following subsections.

\subsection{Model-based change detection framework}
\begin{figure} 
\includegraphics[height= 140pt,width= 250pt]{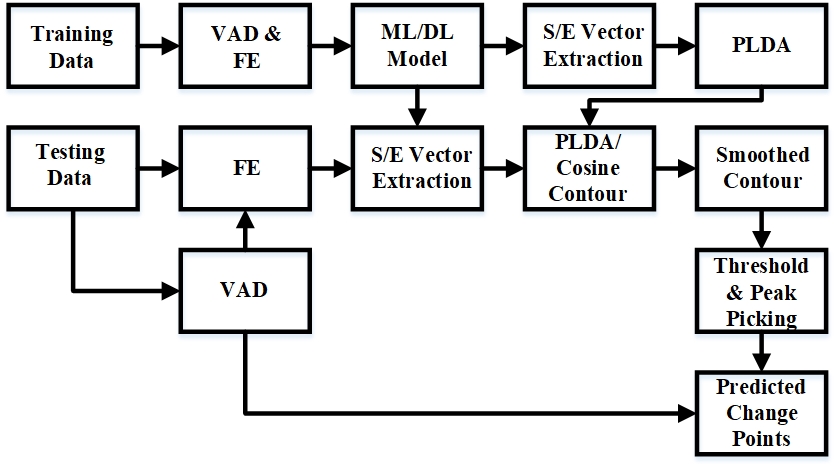}
\caption{Block diagram for the model-based change detection study.}
 \label{mcp}
\end{figure}

The block diagram of the model-based change detection framework is depicted in Fig.~\ref{mcp}. From the training data, initially, MFCC$+\Delta+\Delta\Delta$ are computed, and  voiced feature vectors are selected for further processing by using VAD. The voiced feature vectors are used to train the statistical models like the universal background model (UBM), adaptation model, Total variability matrix (T matrix), and DL  model like TDNN-based x-vector models. The statistical vectors like u/a/i-vectors are extracted using trained UBM/adapt model/ T-matrix, respectively. The u-vector and a-vectors are computed by computing the zeroth order statistics from the UBM and adapt model, respectively. The zeroth order statistics are computed using Equation~\ref{zstat}, where $i$ ranges from $1 \leq i \leq M$, $M$  is the number of mixture components, $x_{j}$ are the MFCC features and $T$ is the number of voiced frames. The u-vectors are the $M$ dimensional vectors extracted using the UBM model, whereas the a-vectors are the concatenation of the $M$ dimensional vectors, extracted from the class-specific adapt models. The i-vectors are extracted as mentioned in~\cite{dehak2010front}. Similarly, the x-vectors are extracted from the trained TDNN-based x-vector model. Both the statistical/ embedding vectors are computed by considering $N$ number voiced feature vectors as analysis window length. The extracted vectors are then used to train the linear discriminate analysis (LDA), within class covariance normalization (WCCN) matrix, and the probabilistic LDA (PLDA) model.

\begin{equation}
\label{zstat}
    N(i)=\frac{1}{T}\sum_{j=1}^{T}P(i|x_{j})
\end{equation}

During testing, the feature vectors are extracted from the code-switched utterances. After that using the VAD labels, with a fixed number of voiced frames the statistical/embedding (S/E) vectors are extracted using the trained models. The S/E vector extraction and the distance contour for each test utterance are computed using Eq.~\ref{dst}. Where $x_{i}$s' are the voiced feature vectors, $\psi(.)$ is the distance computation function and $\mathbb{F}(.)$ is the mapping function from the feature space to S/E vector space. 

\begin{equation}
\label{dst}
    D(i)=\psi(\mathbb{F}(x_{i-N},\ldots,x_{i}),\mathbb{F}(x_{i+1},\ldots,x_{i+N}))
\end{equation}

The distance contour is then smoothed using a hamming window with length ($h_{l}$). The $h_{l}$ is considered as $1/\delta$ times $N$. The peaks of the smoothed contour are computed and the magnitude of peaks greater than the threshold contour is considered as the change points. 
 
\subsection{Experimental Setup}
The TTSF-SC dataset is used for SCD, whereas TTSF-LC and MSCSTB are used for performing LCD tasks. The $39$ dimensional MFCC$+\Delta+\Delta\Delta$ feature vectors are computed from the speech signal with $20$ msec and $10$ msec as window and hop duration, respectively. The voiced frames are decided by considering the frame energy that is greater than the $6\%$ of the utterance's average frame energy. The UBM and adapt models are trained with a cluster size of $32$. The dimensions of the u/a/i-vectors are $32$, $64$ and $50$, respectively. The recipe from the speech brain is used to train and extract the $512$ dimension x-vectors~\cite{ravanelli2021speechbrain}. For the speaker-specific study, the x-vectors are trained without dropout and L2 normalization, whereas for the language-specific study, dropouts of $0.2$ in the second, third, fourth, and sixth layers are used along with L2 normalization.

During training, the speaker/language-specific voiced feature vectors are used to extract the S/E vectors dis-jointly with a fixed $N$, whereas during testing the S/E vectors are extracted with a sample frameshift. All the models have been trained for $20$ epochs. For TTSF-LC the optimal $N$ is decided experimentally as $200$ and for TTSF-SC  $N$ is considered as $50$. After training, by observing the validation loss and accuracy the model corresponding to the $15^{th}$ and $11^{th}$ epoch is chosen for the language and speaker-specific study, respectively. Similarly, for MSCSTB, x-vector models for each language pair are trained. After training for $100$ epochs, by observing the validation loss and accuracy the model belonging to the ($54^{th}$, $29^{th}$, and $26^{th}$) epochs for $N=200$  and ($25^{th}$, $80^{th}$, and $18^{th}$) epochs for $N=50$ are chosen for GUE, TAE, and TEE language pairs, respectively.  

For TTSF-LC and TTSF-SC, the extracted embedding vectors are normalized without having LDA and WCCN. The normalized vectors are used for modeling the PLDA and computing the distance contour for LCD and SCD tasks.  Using the MSCSTB dataset, it is observed that performing LDA, and WCCN along with using cosine kernel distance instead of PLDA distance contour improves the change detection performance. This may be due to the nature of the datasets. The TTSF-LC and TTSF-SC are the studio recording of read speech, whereas the MSCSTB is the conversation recording in the office environment.

For the SCD task, after extracting the i/x vectors, the change points are detected for each test utterance using the hyperparameters $\alpha$, $\delta$, and $\gamma$ as $2.6$, $1.3$ and $0.9$, respectively. The hyper-parameters are decided greedily by observing the change detection performance on the first $100$ test trails. For the LCD task (using TTSF-LC), the hyper-parameters are decided as  $3.2$, $1.3$, and $0.9$, respectively. Similarly, for MSCSTB ($N=200$), the optimal hyperparameters for GUE, TAE, and TEE are ($0.3$, $4.5$, and $1.1$), ($0.3$, $4.5$, and $1.1$) and ($0.3$, $3.9$, and $1.1$), respectively. For $N=50$, the optimal hyperparameters are ($0.3$, $0.9$, and $1.1$), ($0.3$, $0.9$, and $1.3$) and ($0.3$, $0.5$, and $1.3$), respectively.

 \subsection{Language discrimination by statistical/embedding vectors}
 The aim here is to observe the discrimination ability of the extracted S/E vectors for language discrimination, by synthetically emulating the CS scenario. The TTSF-LC, where the same speaker is speaking two languages is considered for this study. The training partition is used to train the UBM, adapt, T-matrix, and TDNN-based x-vector model. From the test partitions, two utterances are selected, one from each language, spoken by a speaker. Using the selected utterances the MFCC$+\Delta+\Delta\Delta$ features and the S/E vectors are extracted and projected in two dimensions using t-SNE~\cite{tsne}.

\begin{figure} 
\includegraphics[height= 190pt,width= 240pt]{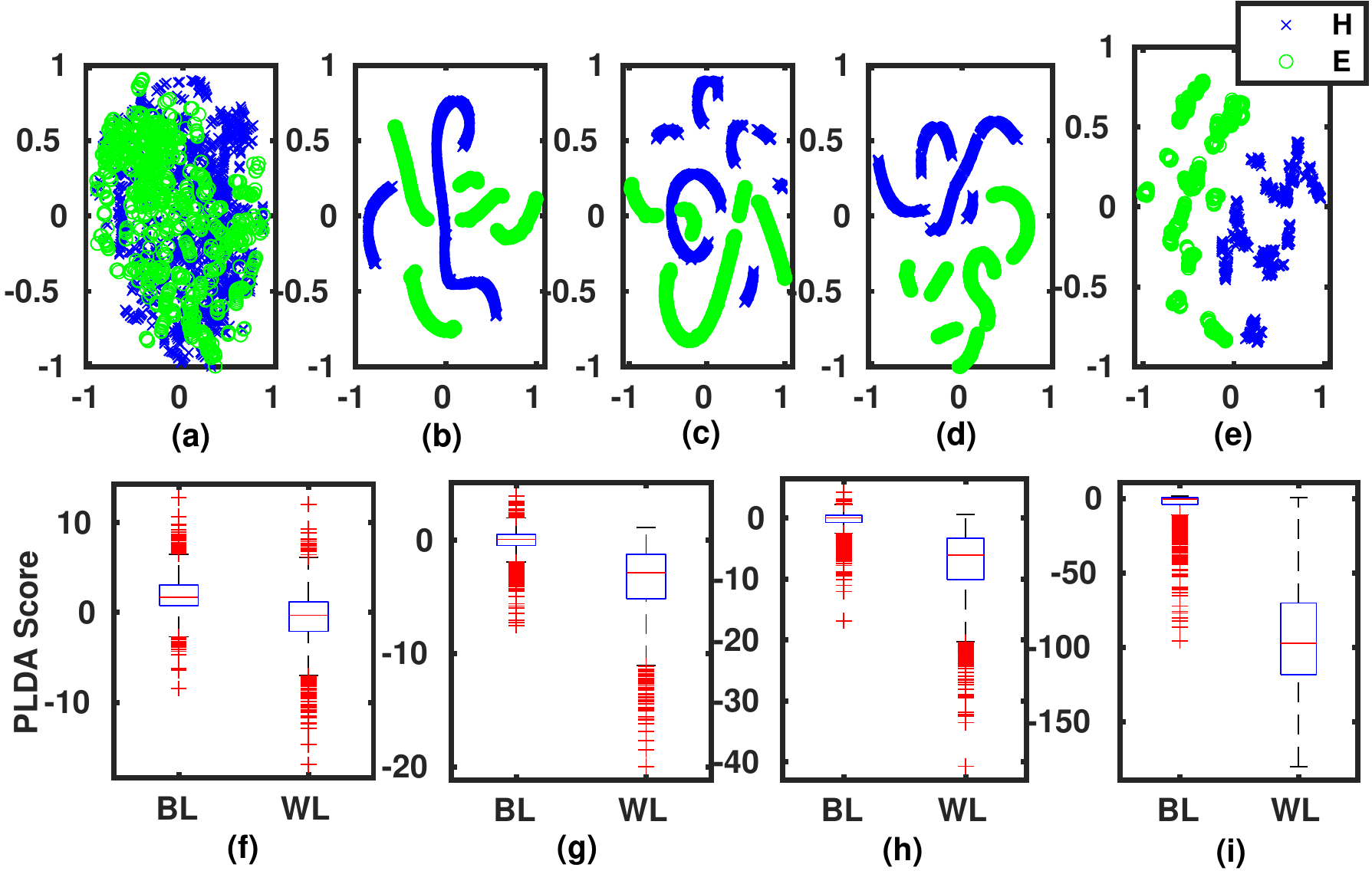}
\caption{t-SNE feature distribution between the Hindi (H) and English (E) (a) MFCC features, (b) u-vector, (c) a-vector, (d) i-vector, and (e) x-vector. Within and Between language PLDA score distribution, with EER of (f) $28.5$, (g) $17.35$,(h) $12.55$, and (i) $3.6$ for u, a, i, and x-vector, respectively.}
 \label{FLD}
\end{figure} 

 
 The two-dimensional representations are depicted in Fig.~\ref{FLD}(a-e). From the figure, it can be observed that the overlapping between the languages reduces by moving from the feature space to the S/E vector space. This shows, like human subjects,  prior exposure to the languages through ML/DL models helps in better discrimination. Furthermore, among the S/E vectors, the overlap between the languages is least in the x-vector space, followed by the i-vector, adapt, and UBM posterior space. This is due to the ability of the modeling techniques to capture the language-specific feature dynamics. 
 
  For strengthening the observation, the features are extracted from the test utterances and pooled together with respect to a given language. The pooled feature vectors are randomly segmented with a context of $200$ and used to extract the S/E vectors. The extracted S/E vectors are paired to form  $2000$ within a language (WL) and $2000$ between language (BL) trails. The WL and BL vector pairs are compared using the PLDA scores. Fig.~\ref{FLD} (f-i) shows boxplots of the PLDA score distribution of the WL and BL pairs. From the box plot distribution, it can be observed that, between the WL and BL, the overlap of PLDA scores distribution reduces with improvement in the modeling techniques from UBM to x-vector.  
 
 In the change point detection task, the aim is to get a sudden change in the distance contour, when there exists a change in language. That can be achieved if the contour (negative of PLDA score) variation is less in WL and provide a sudden change in the contour for BL pairs. Hence for ensuring this, the PLDA score distribution between the WL and BL should be maximized. Keeping this into account, the equal error rate (EER)  has been used as an objective measure, where the WL and BL trials are termed false scores and true scores, respectively. The obtained EER for UBM/adapt/i-vector and x-vector are $28.5$, $17.35$, $12.55$, and $3.6$, respectively. Hence as per the discrimination ability, the change point detection study has been carried out using i/x-vectors as the representations of the speaker and language.

\begin{table*}
\centering
\caption{Performance of LCD and SCD by model-based approaches, S: statistical i-vector, E: embedding based x-vector, N: analysis window length.}
\label{Modtab2}
\begin{tabular}{|c|cc|cc|ccc|ccc|}
\hline
\multirow{3}{*}{} &
  \multicolumn{2}{c|}{\multirow{2}{*}{TTSF-SC}} &
  \multicolumn{2}{c|}{\multirow{2}{*}{TTSF-LC}} &
  \multicolumn{3}{c|}{MSCSTB} &
  \multicolumn{3}{c|}{MSCSTB} \\ \cline{6-11} 
 &
  \multicolumn{2}{c|}{} &
  \multicolumn{2}{c|}{} &
  \multicolumn{1}{c|}{GUE} &
  \multicolumn{1}{c|}{TAE} &
  TEE &
  \multicolumn{1}{c|}{GUE} &
  \multicolumn{1}{c|}{TAE} &
  TEE \\ \cline{2-11} 
 &
  \multicolumn{1}{c|}{S} &
  E &
  \multicolumn{1}{c|}{S} &
  E &
  \multicolumn{3}{c|}{E} &
  \multicolumn{3}{c|}{E} \\ \hline
N &
  \multicolumn{1}{c|}{50} &
  50 &
  \multicolumn{1}{c|}{200} &
  200 &
  \multicolumn{3}{c|}{200} &
  \multicolumn{3}{c|}{50} \\ \hline
IDR &
  \multicolumn{1}{c|}{87.75} &
  92.27 &
  \multicolumn{1}{c|}{80.58} &
  87.01 &
  \multicolumn{1}{c|}{46.56} &
  \multicolumn{1}{c|}{49.91} &
  47.13 &
  \multicolumn{1}{c|}{54.74} &
  \multicolumn{1}{c|}{52.19} &
  50.84 \\ \hline
FAR &
  \multicolumn{1}{c|}{5.42} &
  3.96 &
  \multicolumn{1}{c|}{8.8} &
  8.84 &
  \multicolumn{1}{c|}{5.95} &
  \multicolumn{1}{c|}{10.42} &
  6.50 &
  \multicolumn{1}{c|}{13.10} &
  \multicolumn{1}{c|}{27.82} &
  19.34 \\ \hline
MDR &
  \multicolumn{1}{c|}{6.83} &
  3.76 &
  \multicolumn{1}{c|}{10.57} &
  4.41 &
  \multicolumn{1}{c|}{47.49} &
  \multicolumn{1}{c|}{39.67} &
  46.36 &
  \multicolumn{1}{c|}{32.16} &
  \multicolumn{1}{c|}{19.99} &
  29.83 \\ \hline
Dm &
  \multicolumn{1}{c|}{0.05} &
  0.03 &
  \multicolumn{1}{c|}{0.33} &
  0.28 &
  \multicolumn{1}{c|}{0.51} &
  \multicolumn{1}{c|}{0.56} &
  0.56 &
  \multicolumn{1}{c|}{0.35} &
  \multicolumn{1}{c|}{0.30} &
  0.34 \\ \hline
\end{tabular}
\end{table*}

 \subsection{Experimental Results}
 
Initially, the change detection study is conducted with TTSF-SC and TTSF-LC  using i/x-vectors as the speaker/language representation. The discrimination ability and the LCD/SCD study suggest that the x-vector is a better representation of the speaker/language than the i-vector. Therefore, the LCD task on the MSCSTB dataset is conducted by considering x-vectors as language representations.  

 The experimental results are tabulated in Table~\ref{Modtab2}. The performance obtained in terms of IDR on SCD task using i-vector and x-vector is $87.75\%$ and $92.27\%$, respectively. Similarly, for LCD tasks the performances on TTSF-LC are $80.58\%$ and $87.01\%$, respectively. As evidenced by the language discrimination study, the performance of LCD provides a relative improvement of $21.9\%$ and $31.63\%$ using i-vectors and x-vectors, over the best performance achieved on the unsupervised distance-based approach, respectively. This justifies the claim that, like humans, the performance of the LCD can be improved by incorporating language-specific prior information through computational models.

 \begin{figure} 
\includegraphics[height= 110pt,width= 240pt]{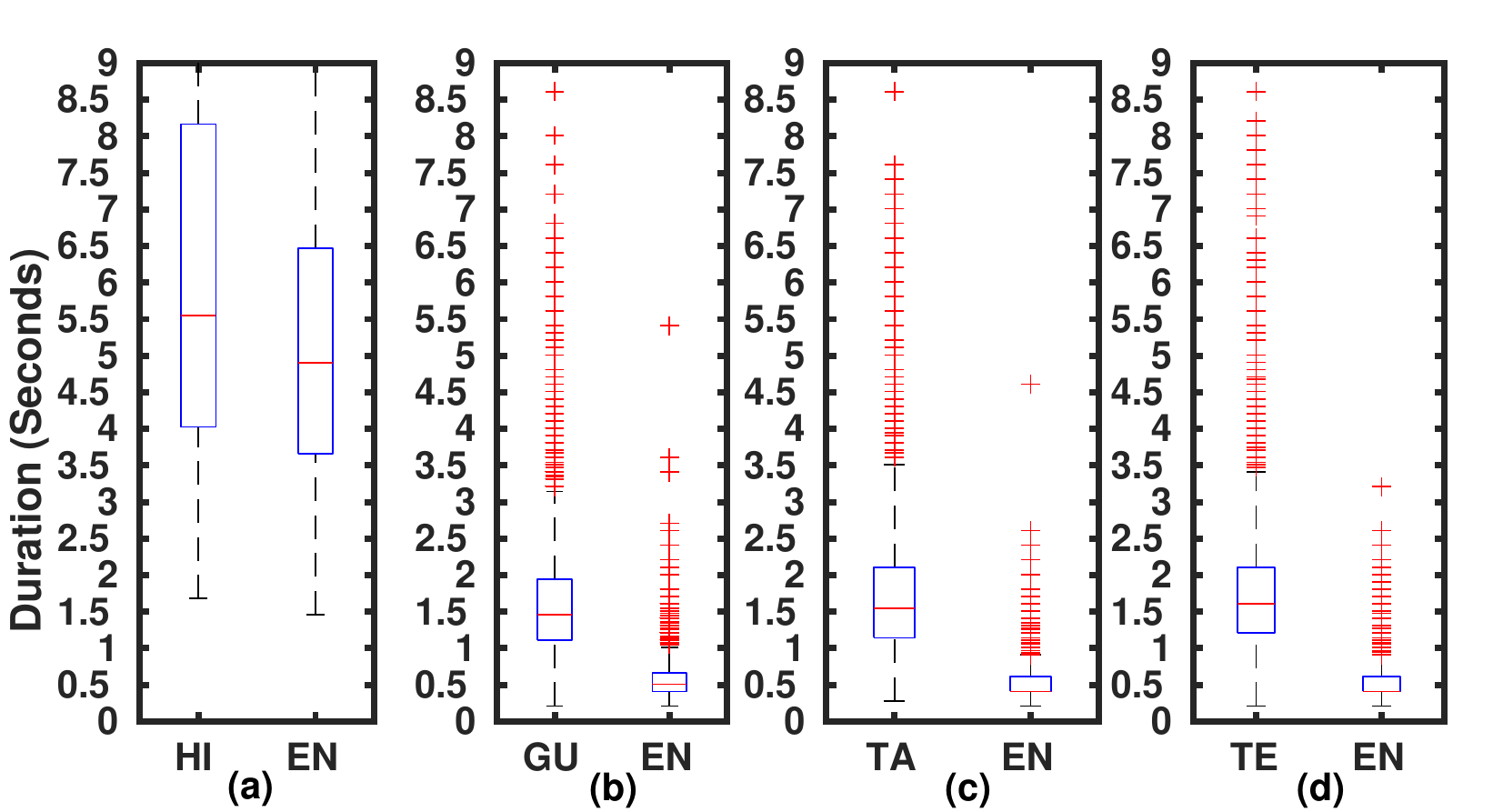}
\caption{ Distribution of the mono-lingual segment duration of (a) TTSF-LC's HE, and MSCSTB's  (b) GUE, (c) TAE, and (d) TEE language pairs, respectively.}
 \label{dur}
\end{figure} 

The performance of the LCD task on the MSCSTB dataset using x-vectors as language representation with considering N as $200$ (same as TTSF-LC) is $46.56\%$, $49.91\%$ and $47.13\%$ in terms of IDR for the GUE, TAE, and TEE partitions, respectively. The performance provides a relative improvement of $5.6\%$, $2.3\%$, and $4.2\%$. However, the improvement is small as compared to the improvement achieved using TTSF-LC data. This may be due to the distributional difference in the monolingual segment duration in the TTSF-LC and MSCSTB datasets. A boxplot showing the distribution of the monolingual segments of TTSF-LC's and MSCSTB's test set is depicted in Fig. ~\ref{dur}. From the figure, it can be observed that the median of the monolingual segment duration in the case of TTSF-LC for primary and secondary language are ($5.54$ and $4.9$) seconds, and for MSCSTB is ($1.46$ and $0.51$), ($1.54$ and $0.41$), ($1.61$ and $0.41$) seconds for GUE, TAE, and TEE partition, respectively. Further, it has been observed that language discrimination is better by considering $N$ equal to $200$ (i.e. approx. $2$ seconds). Hence, due to the monolingual segment duration of the MSCSTB dataset being smaller than the considered analysis window duration resulting in smoothing on the resultant distance contour, and leads to an increase in the MDR. Therefore, the alternative is to reduce the analysis window length, but that may affect the language discrimination ability of the x-vectors.

\begin{table}[]
\centering
\caption{Perforance of LCD by varying the analysis window length.}
\label{gue_lcd}
\begin{tabular}{|c|ccccc|}
\hline
\multirow{2}{*}{} & \multicolumn{5}{c|}{MSCSTB}                                                                                               \\ \cline{2-6} 
                  & \multicolumn{5}{c|}{GUE}                                                                                                  \\ \hline
N                 & \multicolumn{1}{c|}{200}   & \multicolumn{1}{c|}{150}   & \multicolumn{1}{c|}{100}   & \multicolumn{1}{c|}{75}    & 50    \\ \hline
IDR               & \multicolumn{1}{c|}{46.56} & \multicolumn{1}{c|}{48.12} & \multicolumn{1}{c|}{50.12} & \multicolumn{1}{c|}{51.78} & 54.74 \\ \hline
FAR               & \multicolumn{1}{c|}{5.95}  & \multicolumn{1}{c|}{7.38}  & \multicolumn{1}{c|}{12.54} & \multicolumn{1}{c|}{11.01} & 13.10 \\ \hline
MDR               & \multicolumn{1}{c|}{47.49} & \multicolumn{1}{c|}{44.50} & \multicolumn{1}{c|}{37.35} & \multicolumn{1}{c|}{37.21} & 32.16 \\ \hline
Dm                & \multicolumn{1}{c|}{0.51}  & \multicolumn{1}{c|}{0.48}  & \multicolumn{1}{c|}{0.37}  & \multicolumn{1}{c|}{0.33}  & 0.35  \\ \hline
\end{tabular}
\end{table}

\begin{figure} 
\includegraphics[height= 110pt,width= 240pt]{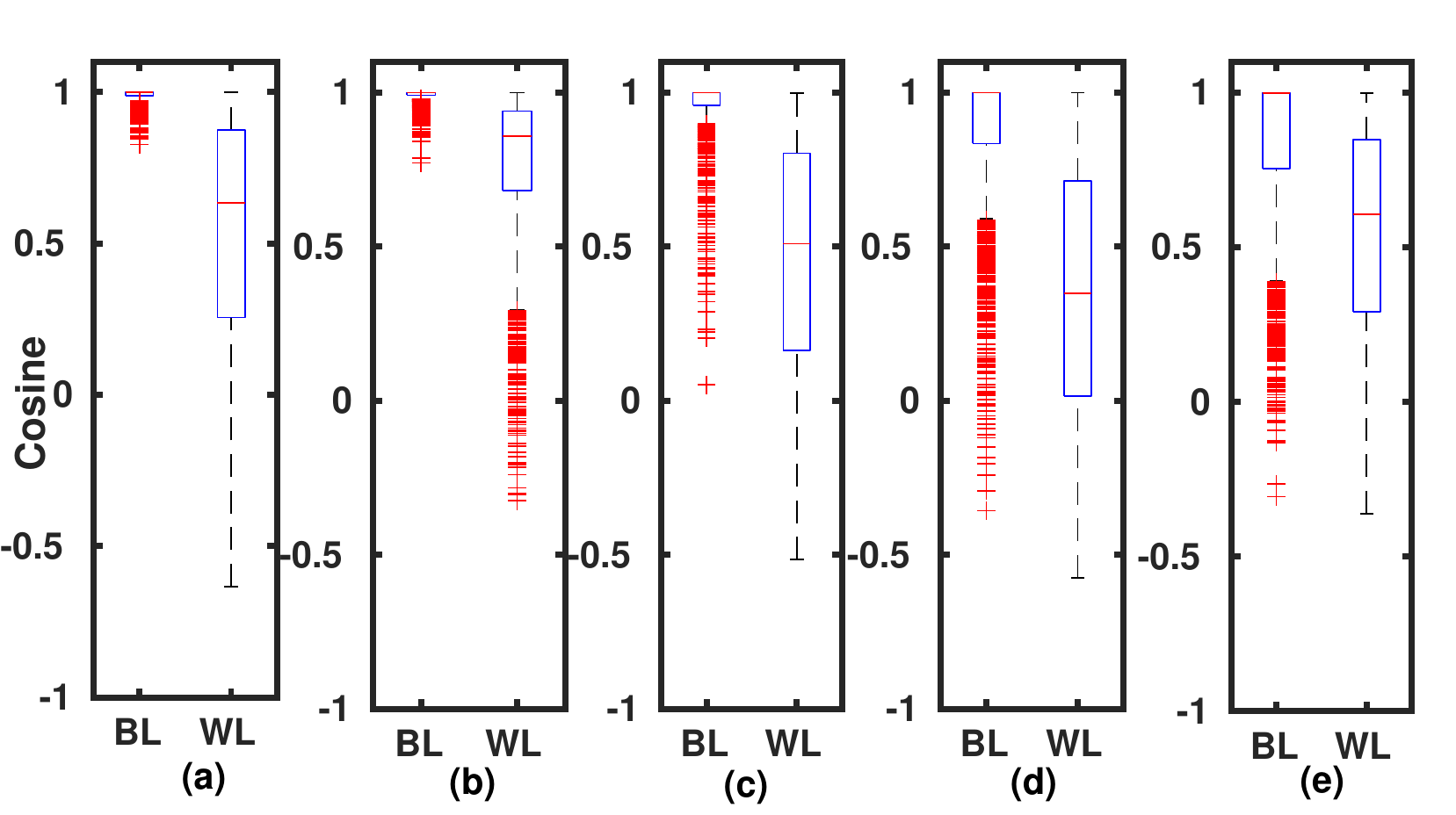}
\caption{ Within and Between language cosine score distribution, with EER of (a) $7.1$, (b) $9.8$,(c) $12.8$, (d) $19.8$ and (i) $29.2$ for the analysis window length of $200$, $150$, $100$, $75$ and $50$, respectively.}
 \label{dis_dur}
\end{figure}

A study is performed for observing the trade-off between the analysis window length and language discrimination ability. The language discrimination test and the LCD task are performed using the GUE partition of the MSCSTB dataset by reducing the analysis window length from $200$ to $50$. The results of the LCD task are tabulated in Table~\ref{gue_lcd}. The cosine score distribution of the x-vectors'  WL and BL pairs after the LDA and WCCN projection with varying the analysis window length are depicted in Fig.~\ref{dis_dur}. From the Table, it can be observed that with decreasing in $N$, the performance of the LCD task improves, and achieved the best performance of $54.74\%$  at $N$ equals to $50$. Hence the change detection performance is computed with $N$ equal to $50$ for GUE, TAE, and TEE language pairs and tabulated in Table~\ref{Modtab2}. However, the relative performance improvement by incorporating language-specific prior exposure through the x-vector model is not as expected as in the TTSF-LC dataset. This is due to the language discrimination ability of the x-vectors reducing with the decrease in $N$. From Fig.~\ref{dis_dur}, it can be observed that the overlap between the WL and BL score distribution increases with a decrease in the value of $N$. As an objective measure, the computed EER for $N$ equals to $200$, $150$, $100$, $75$ ,and $50$ are $7.1$, $9.8$, $12.8$, $19.8$ and $29.2$, respectively.

\section{Discussion}
The human-based LCD and SCD study suggests that the language requires more neighborhood information as compared to the speaker for comfortable discrimination. Further, prior exposure to the languages helps humans to better discriminate between the languages. Motivated by this, it is hypothesized that the performance of LCD by machine can be improved with the 
 (a) incorporation larger duration analysis window ($N$) and (b) language-specific exposure through computational models. 

In the unsupervised distance-based approach, it has been observed that the performance of the LCD improves by increasing the value of $N$. The optimal $N$ value for the SCD study is $150$. Considering the same value of $N$, the LCD task is carried out for both TTSF-LC and MSCSTB datasets, and performances are tabulated in Table~\ref{pcom}. In the case of the MSCSTB dataset, the average IDR values with respect to all three language pairs are tabulated. Motivating by the LCD/SCD study by humans, the $N$ value is increased and the obtained optimal $N$ value for the LCD with TTSF-LC is $200$. Similarly, the optimum N value for MSCSTB  is $160$, $180$, and $170$ for the GUE, TAE, and TEE , respectively. The performance with the optimal $N$ value for TTSF-LC and MSCSTB is $66.1\%$ and $46.02\%$, which provides a relative improvement of $29.1\%$, and $2.4\%$, respectively. These observations justify the claim that the performance of the LCD by machines can be improved by increasing the analysis window duration.

\begin{table}[]
\centering
\caption{Performance comparison, RI: relative improvement, A: with $N=150$ (tuned for SCD), B: with the optimal $N$ value (tuned for LCD), and C: x-vector based approach.}
\label{pcom}
\begin{tabular}{|c|c|c|c|}
\hline
Dataset                  & Approach & IDR   & RI    \\ \hline
\multirow{2}{*}{TTSF-SC} & A        & 84.1  & -     \\ \cline{2-4} 
                         & C        & 92.27 & 9.71  \\ \hline
\multirow{3}{*}{TTSF-LC} & A        & 51.2  & -     \\ \cline{2-4} 
                         & B        & 66.1  & 29.1  \\ \cline{2-4} 
                         & C        & 87.01 & 31.63 \\ \hline
\multirow{3}{*}{MSCSTB}  & A        & 44.93 & -     \\ \cline{2-4} 
                         & B        & 46.02 & 2.4   \\ \cline{2-4} 
                         & C        & 52.59 & 14.27 \\ \hline
\end{tabular}
\end{table}

Furthermore, as hypothesized from the subjective study, the incorporation of language-specific exposure through computational models improves LCD performance. The i/x-vector models have been trained, which essentially capture the language-specific cepstral dynamics. It has been observed that with the x-vector approach, the obtained performance is $87.01\%$ for TTSF-LC and $52.59\%$  in terms of IDR, which provides a relative improvement of $31.63\%$ and $14.27\%$ over the performance of the unsupervised distance-based approach. Similarly, for the SCD task using the TTSF-SC dataset, the performance provides a relative improvement of $9.71\%$. Comparing the performance of LCD and SCD on synthetic data, it can be observed that the improvement is more significant on LCD than the SCD. This concludes, like human subjective study, in an ideal condition (only speaker/language variation and keeping other variations limited), the requirement model-based approach is more significant on LCD than the SCD.  

It is also observed that in the LCD task, the performance improvement on MSCSTB data is limited as compared to the improvement achieved on the synthetic TTSF-LC dataset. This is due to the difference in the mono-lingual segment duration. The trade-off between the analysis window duration and the language discrimination ability shows that the discrimination ability improves with an increase in analysis window duration. At the same time during change detection, as the mono-lingual segment duration can possibly be lesser than $500$ msec (approx. $50$ voiced frames), considering a larger analysis window leads to degrading in performance by smoothening the evidence contour (leads to an increase in MDR). Hence to overcome this issue, (1) need to achieve significant language discrimination with the $N$ value as small as possible, and (2) need to develop a framework whose performance will be least affected/independent with the variations of the analysis window duration.


\section{Conclusion}
\label{sec:5}
In this work, we performed LCD using the available frameworks for SCD.  From the subjective study, it is observed that humans require comparatively larger neighborhood information around the change point as compared to the speaker. It is also observed that prior language-specific  exposure improves the performance of the LCD task. In the unsupervised distance-based approach, the incorporation of larger neighborhood information improves the LCD performance by relatively $29.1\%$ and $2.4\%$ on the synthetic TTSF-LC and the practical MSCSTB dataset, respectively. Similarly, incorporating language-specific prior information through the computational models provides a relative improvement of $31.63\%$ and $14.27\%$ over the unsupervised distance-based approach.

It has also been observed that the practical data set does not perform as expected like synthetic data. This is due to the distributional difference in the monolingual segment duration on both datasets. The MSCSTB dataset consists of the monolingual segments having a duration lesser than $0.5$ secs, and for better language discrimination the required duration is about $2$ secs (about $200$ voiced frames). Hence it is challenging to decide on the analysis window duration. The larger duration smooths the evidence contour and increases the MDR, whereas a smaller duration of $0.5$ secs is not able to provide appropriate language discrimination. 

Therefore, our future attempts will try to develop a better framework, which can provide better language discrimination on a small duration, and also plan to come up with a change detection framework, whose performance should be independent/less affected by the variations of the analysis window duration.


 
\begin{acknowledgments}
The authors like to acknowledge "Anatganak", high-performance computation (HPC) facility, IIT Dharwad, for enabling us to perform our experiments. And the Ministry of Electronics and Information Technology (MeitY), Govt. of India, for supporting us through different projects.
\end{acknowledgments}

\bibliography{sampbib}
\end{document}